\documentclass{emulateapj}

\usepackage{rotating}
\usepackage[caption=false]{subfig}

\slugcomment{}

\shorttitle{Extragalactic SETI}
\shortauthors{Zackrisson et al.}

\begin{document}
\title{Extragalactic SETI: The Tully-Fisher relation as a probe of Dysonian astroengineering in disk galaxies}
\author{Erik Zackrisson\altaffilmark{1,2}$^*$, Per Calissendorff\altaffilmark{2}, Saghar Asadi\altaffilmark{2} \& Anders Nyholm\altaffilmark{2}}
\altaffiltext{*}{E-mail: erik.zackrisson@physics.uu.se}
\altaffiltext{1}{Department of Physics and Astronomy, Uppsala University, Box 515, SE-751 20 Uppsala, Sweden}
\altaffiltext{2}{Department of Astronomy, AlbaNova, Stockholm University, SE-106 91 Stockholm, Sweden}

\begin{abstract}
If advanced extraterrestrial civilizations choose to construct vast numbers of Dyson spheres to harvest radiation energy, this could affect the characteristics of their host galaxies. Potential signatures of such astroengineering projects include reduced optical luminosity, boosted infrared luminosity and morphological anomalies. Here, we apply a technique pioneered by \citet{Annis} to search for Kardashev type III civilizations in disk galaxies, based on the predicted offset of these galaxies from the optical Tully-Fisher relation. By analyzing a sample of $ 1359$ disk galaxies, we are able to set a conservative upper limit of $\lesssim 3\%$ on the fraction of local disks subject to Dysonian astroengineering on galaxy-wide scales. However, the available data suggests that a small subset of disk galaxies actually may be underluminous with respect to the Tully-Fisher relation in the way expected for Kardashev type III objects. Based on the optical morphologies and infrared-to-optical luminosity ratios of such galaxies in our sample, we conclude that none of them stand out as strong Kardashev type III candidates and that their inferred properties likely have mundane explanations. This allows us to set a tentative upper limit at $\lesssim 0.3\%$ on the fraction of Karashev type III disk galaxies in the local Universe. 
\end{abstract}





\section{Introduction}
\label{intro}
For more than 50 years, astronomers have searched the skies for artificial signals and other signs of extraterrestrial civilizations \citep[e.g.][]{Tarter a,Tarter b}, yet no evidence of intelligent life beyond the Earth has so far emerged. One viable explanation, among the many proposed for this so-called Fermi paradox \citep[e.g.][]{Brin,Webb,Cirkovic} is that technologically advanced civilizations are very rare. Most SETI efforts have so far been limited to the confines of the Milky Way, and with $\gtrsim 10^{11}$ large galaxies in the observable Universe \citep{Beckwith et al.}, it has been argued that the prospects of detection may be better if the search radius is extended to extragalactic scales \citep[e.g.][]{Cirkovic & Bradbury,Wiley}. The transmission of signals across extragalactic distances  would be far more challenging than across a single galaxy (see e.g. \citealt{Maccone} for a proposed technique of this type), but extremely advanced supercivilizations (Kardashev type II-III; \citealt{Kardashev}) could in principle be detectable through indirect signatures of galactic-scale astroengineering projects even if they have no interest in intentionally transmitting signals in our direction \citep{Annis,Carrigan 10,Inoue & Yokoo,Wright et al. b,Wright et al. c,Griffith et al.}. 

One potential way to harvest the radiation energy of a star is the construction of a Dyson sphere \citep{Dyson}, which would either completely or partially enshroud the star and make a significant fraction of radiation energy available for supercomputing or other project with immense power requirements \citep[e.g.][]{Sandberg,Bradbury}. This would render the star unusually faint or completely dark at ultraviolet/optical wavelengths, but could make it bright at infrared wavelengths due to the need to get rid of excess heat. Several unsuccessful searches for individual Dyson spheres within the Milky Way have already been carried out \citep{Slysh, Timofeev et al.,Jugaku & Nishimura,Carrigan 09}. 

In a pioneering study, \citet{Annis} proposed the use of galaxy scaling relations to single out potential Kardashev type III candidates, i.e. galaxies in which an extremely advanced civilization has spread throughout its host galaxy and converted vast numbers of stars into Dyson spheres. As such galaxy-scale astroengineering projects would likely not affect the gravitational potential of the galaxy, but substantially decrease the total optical luminosity, host galaxies of Kardashev type III civilizations may appear anomalously underluminous for their type. In a sample of 57 disk galaxies and 106 ellipticals, \citet{Annis} detected no significant outliers. Here, we revisit the \citet{Annis} method and apply it to a Tully-Fisher sample of 1359 disk galaxies. 

The search method, the detected Tully-Fisher outliers and the resulting upper limits on the prevalence of star-fed Kardashev type III civilizations in disk galaxies are discussed in Sect.~\ref{constraints}. In Sect.~\ref{nature}, the nature of the outliers is probed further using morphological information and infrared data. Other potential signatures of supercivilizations that make use of Dyson spheres on large scales are discussed in Sect.~\ref{discussion}. Sect.~\ref{summary} summarizes our findings.

\section{Kardashev type III candidates in the Tully-Fisher diagram}
\label{constraints}
The Tully-Fisher relation \citep{Tully & Fisher} is an empirical relation between the velocity width (amplitude of the rotation curve) of a spiral galaxy and its absolute magnitude. Because of the low scatter in this relation, it is often used as a distance indicator: By combining an apparent magnitude derived from photometry with a spectroscopic measurement of the velocity width (usually based on the HI 21 cm line or the H$\alpha$ line at 6563 \AA), it becomes possible to infer the distance by adopting the absolute magnitude predicted by the relation. The Tully-Fisher relation has been shown to hold for rotationally supported galaxies over an impression range in surface brightness and luminosity \citep[e.g.][]{Zwaan et al.,McGaugh98,Chung et al.}, although there are indications of increased scatter and a systematic change in slope for the faintest and smallest galaxies \citep{Matthews98,Geha et al.}. 

As pointed out by \citet{Annis}, galaxy-wide colonization projects in which large numbers of stars are converted into Dyson spheres would affect the absolute magnitude in short-wavelength passbands, but are not likely to affect the velocity width, which serves as a proxy for the depth of the gravitational well. The host galaxies of star-fed Kardashev type III (hereafter KIII) civilizations should therefore appear significantly underluminous compared to the optical Tully-Fisher (hereafter TF) relation. In search of KIII host galaxy candidates, \citet{Annis} adopted the criterion that such objects should have their optical luminosity dimmed by at least a factor of four (i.e. 1.5 magnitudes) compared to the TF relation, and analyzed a sample of 57 disk galaxies from \citet{Pierce & Tully} without finding any TF outliers according to this criterion. Here, we apply the same technique to TF data (luminosity in the $I$-band and velocity width $W_\mathrm{TF}$ based on either the HI 21 cm line or H$\alpha$) for $ 1359$ objects from the \citet{Springob et al. 07} SFI++ catalog. 

The criterion that KIII galaxy candidates should be located $\geq 1.5$ mag below the TF is simply chosen for convenience -- the TF relation has an intrinsic scatter \citep[e.g.][]{Masters et al.} that, when coupled to observational errors would lead to too many spurious detections if the limit had been placed closer to the TF. The baryonic TF \citep[][]{McGaugh00} exhibits smaller intrinsic scatter and could potentially be used to search for less extreme KIII objects, but this requires additional data for each object (an estimate of the molecular gas fraction), and current baryonic TF samples are consequently much smaller. Objects located significantly more than 1.5 mag below the TF become increasingly likely to be excluded from existing TF samples due to selection effects (which we attempt to quantify in Sect.~\ref{statistics}). For these reasons, we have chosen to stick to the $\geq 1.5$ mag criterion used by \citet{Annis}. Using the PARSEC v1.2S isochrones \citep{Bressan12,Chen14,Tang14}, as implemented in the CMD\footnote{http://stev.oapd.inaf.it/cmd} spectral synthesis code, we estimate that 40-50\% of the stellar contribution to the $I$-band flux of galaxies that have undergone active star formation for $\sim 10$ Gyr comes from main-sequence stars, whereas the remaining fraction comes from more evolved stars (red giants, horizontal branch stars, asymptotic giant stars). Hence, a very wide range of stars would need to be enshrouded in close-to-complete Dyson spheres in order for KIII civilizations to be detectable with the TF method.

Below, we present a brief overview of how the TF data in the SFI++ catalog was originally derived. Further pruning and modifications of this catalog for the purposes of our KIII search are described in Sect.~\ref{peculiar}.

In the \citet{Springob et al. 07} data set, $W_\mathrm{TF}$ is either derived from spectroscopy of the HI 21 cm line or from optical H$\alpha$ rotation curves.  In the HI case (the most common case for the TF outliers we find), Springob et al derive the line width $W_\mathrm{HI}$ using:
\begin{equation}
W_\mathrm{TF}=\left( \frac{W_\mathrm{HI, obs}-\Delta_\mathrm{ins}}{1+z} -\Delta_\mathrm{t} \right)\frac{1}{\sin i}
\label{W21_eq}
\end{equation}
where $W_\mathrm{HI, obs}$ represents the observed line width of the HI 21 cm line, $z$ the redshift (based on the wavelength of either the HI 21 cm line or the wavelength of H$\alpha$), $i$ the inclination, $\Delta_\mathrm{ins}$ the instrumental correction to the line width and $\Delta_\mathrm{t}$ the corresponding turbulence correction. In the case of $\Delta_\mathrm{t}$, a value of 6.5 km s$^{-1}$ was adopted. For a detailed description of the instrumental line width correction $\Delta_\mathrm{t}$, see \citet{Springob et al. 05}.

In the case when optical rotation curves are used in SFI++, corrections are applied for known systematic differences between H$\alpha$ and HI velocity widths due to the limited extent of the H$\alpha$ data. The inclinations used to correct $W_\mathrm{TF}$ for orientation effects in the catalog is derived using
\begin{equation}
i = \cos^{-1} \left( \frac{(b/a)^2 - q_0^2}{1-q_0} \right)^{0.5}.
\label{inclination_eq}
\end{equation}
Here, $a$ represents the semimajor axis, $b$ the semiminor axis, $q_0$ the intrinsic axial ratio $b/a$ of the disk at edge-on orientation, assumed to be $q_0=0.13$ for disks of morphological type Sbc and later, and 0.20 for disks of earlier morphological types.

The $I$-band absolute magnitudes of SFI++ have been corrected for internal dust attenuation by an amount $\Delta M_\mathrm{att}$ given by
\begin{equation}
\Delta M_\mathrm{att} = -\gamma \log_{10} (a/b),
\label{extinction_eq}
\end{equation}
where $\gamma$ is taken to be a function of $M_I$ (with high-luminosity disks assumed to display higher dust attenuation; see \citealt{Springob et al. 07} for details). The data have also been corrected for Galactic extinction using COBE data \citep{Schlegel et al.}. 

Pinning down the factors that dominate the error budget on the position of an object in the TF diagram is highly nontrivial, given the many measurements and corrections involved, but \citet{Giovanelli et al. b} attempt such an analysis for a subset of the SFI++ data. They find that the errors and corrections on the velocity width tend to dominate, except in the case of high-width, high-luminosity, highly inclined galaxies, for which errors on the extinction correction (eq.~\ref{extinction_eq}) can become the main source of uncertainty. The median correction for dust in our sample is $\approx 0.3$ mag, but in extreme cases, the correction does reach $\approx 1.5$ mag. Errors on photometry, redshift and ellipticity are typically negligble by comparison. Of course, this holds for normal galaxies, absent KIII activity. The sensitivity of the dust correction to the optical axial ratio $(b/a)$ could in principle jeopardize the assumption that Dysonian astroengineering should affect the optical stellar luminosity (as measured by $M_I$), while leaving the overall dynamics and the gas phase (as measured by  $W_\mathrm{TF}$) untouched. If KIII activity results in morphological curiosities which affect the apparent $(b/a)$ ratio, then this could affect both $M_I$ and $W_\mathrm{TF}$, through the inclination-related corrections going into both eq.(\ref{W21_eq}) and eq.(\ref{extinction_eq}). 

Based on the numerical simulations of KIII colonization presented in Sect.~\ref{simulations}, we argue that this effect can go both ways, and act to either accentuate or mask KIII activity in the TF diagram. An evenly colonized KIII disk with uniformly diminished surface brightness may appear rounder than it actually is (due to a spuriously high $b/a$), since only the inner, bulge-dominated parts may remain visible above the surface brightness threshold in optical images. Such objects would be pushed towards the lower right of the TF diagram (deeper into the KIII candidate region), since $W_\mathrm{TF}$ would be biased high and the dust correction on $M_I$ would be underestimated. An already highly inclined galaxy in which KIII colonization has focused on the bulge-dominated inner regions may on the other hand appear flatter than it actually is (due to a spuriously low $b/a$), thereby pushing it upwards (towards the TF line and away from the KIII region). This sensitivity to how KIII colonization has progressed within a galaxy is an inherent limitation of the TF approach to KIII searches, as long as the axial ratio is based on the optical appearance, and should be kept in mind as a possible caveat when considering the TF-based limits on astroengineered disks we derive in Sect.~\ref{statistics}.

\subsection{Distance estimates}
\label{peculiar}
In the original SFI++ catalog, distances (in velocity units) to galaxies not belonging to any group or cluster are estimated using the relation
\begin{equation}
r_{\mathrm{gal}} = cz_{\mathrm{gal}} - v_{\mathrm{gal}},
\label{distance_eq}
\end{equation}
where $cz_{\mathrm{gal}}$ is the velocity in the CMB frame and  $v_{\mathrm{gal}}$ the peculiar velocity expressed as
\begin{equation}
v_{\mathrm{gal}}  = cz_{\mathrm{gal}} (1-10^{0.2 dm}).
\label{velocity_eq}
\end{equation}
Here, $dm$ represents the difference between the attenuation-corrected absolute magnitude (as presented in the SFI++ catalog) and the predicted absolute magnitude one would expect if the object had no offset from the TF relation. 

However, since we are specifically looking for TF outliers, distances based on the assumption that a single object lies on the TF are of little use. For objects belonging to galaxy groups or clusters containing several disk galaxies with individual TF distances, the average distance to the group or cluster itself may still be used as an estimate of the distance to the member galaxies. 

In our analysis, we therefore discard all field galaxies from the sample and estimate new absolute magnitudes $M_I$ for the remaining objects based on the SFI++ distances assigned to the groups and clusters to which they are deemed to belong.

In \citet{Springob et al. 07}, such group and cluster distances are estimated similar to eq.~(\ref{distance_eq}), but with $cz_{\mathrm{grp}}$ calculated as the mean redshift of all group members and peculiar velocities estimated as
\begin{equation}
v_{\mathrm{grp}} = \frac{\sum_{i=1, N_{\mathrm{SFI}++}} v_i/\epsilon_i}{\sum_{i=1, N_{\mathrm{SFI}++}} 1/\epsilon_i},
\label{groupvelocity_eq}
\end{equation}
where $v_i$ is calculated similar to eq.~(\ref{velocity_eq}), with the slight difference of using $cz_{\mathrm{grp}}$ instead of $cz_{\mathrm{gal}}$ for individual galaxies. Here $\epsilon_i$ represents the estimated error on $v_i$. \citet{Springob et al. 07} also present a version of $v_{\mathrm{grp}}$ corrected for Malmqvist bias, $r_\mathrm{grp,M}$ (distance in velocity units), and this is the quantity we use to recompute the absolute magnitudes: 
\begin{equation}
M_I = m_{I\mathrm{c}} -5 \log_{10} (r_{\mathrm{grp,M}}/H_0) +5.
\label{distance_modulus}
\end{equation}
Here, $m_{I\mathrm{c}}$ is the $I$-band apparent magnitude corrected for both internal and Galactic extinction and $H_0$ is the Hubble constant. To be consistent with the luminosity-dependent extinction correction scheme derived by \citet{Giovanelli et al.} and adopted by \citet{Springob et al. 07}, we have adopted $H_0=100$ km s$^{-1}$ Mpc$^{-1}$ when computing the extinction corrections, but have then rescaled all absolute magnitudes to $H_0=70$ km s$^{-1}$ Mpc$^{-1}$, which is more in line with modern measurements of the Hubble parameter. One should note that the extinction correction scheme derived by \citet{Giovanelli et al.} was based mainly on Sbc and Sc galaxies, whereas SFI++ contains a greater diversity of spiral subclasses. 
\citet{Springob et al. 07} argue that, because of this, extinction in earlier types may possibly be underestimated. However, this turns out to be of little consequence for the present study, since very few Sa galaxies end up as KIII candidates (see Sect.~\ref{nature} for more discussion on Sa subtypes).

The underlying assumption when giving preference to group distances over the distances to individual objects given in SFI++ is that star-fed KIII civilizations are relatively rare (as already indicated by the studies of \citealt{Annis} and \citealt{Wright et al. c}), so that even though a single group/cluster member may be a TF outlier, the TF distance to the group/cluster as a whole need not be significantly biased. In the case of a colonization wave spreading across intergalactic distances from one galaxy to the next, this assumption may of course be jeopardized, as further discussed in Sect.~\ref{discussion}. A second caveat is that, since groups/clusters can have non-negligible extent along the line of sight, the distance to a member galaxy may be somewhat different from the distance to the group/cluster centre. 

After rejecting field galaxies from SFI++, we further prune the sample by removing galaxies belonging to groups and clusters with less than 3 SFI++ members, since too large distance errors would otherwise be introuced (see discussion in Sect.~\ref{reliability}). Out of 4861 galaxies in the full SFI++ catalog, our remaining subsample consists of 1359 objects. Our scheme for reassessing absolute magnitudes for group and cluster members then generates a sample of 11 galaxies seemingly located $\geq 1.5$ mag below the TF relation, which thereby qualify as KIII candidates according to the \citet{Annis} criterion. The properties of these galaxies are summarized in Table~\ref{KIII_candidates}.  

As it turns out, our 11 KIII candidates all belong to the so-called nontemplate groups of the SFI++ catalog, i.e. groups of galaxies with relatively few other SFI++ members \citep{Masters et al.,Springob et al. 07}. As the exact spatial sizes of the groups defined by \citet{Springob et al. 07} are not known, we introduce an uncertainty on the absolute magnitudes by assigning each group an estimated radius of 1.5 Mpc. Assuming this to be the outer edge that confines the group, we calculate the absolute magnitude that our objects would have at this distance and adopt the difference between the central and edge magnitude as the uncertainty. As a result, the uncertainty in absolute magnitude consequently grows with decreasing group distance (reaching a maximum of $\pm 0.25$ mag for the most nearby object in Table~\ref{KIII_candidates}). In order to include uncertainties from extinction and inclination we also add the error of the absolute magnitude tabulated by \citet{Springob et al. 07} to the total error budget. 

\subsection{Quality of the KIII candidates}
\label{reliability}
One of the risks involved in deriving absolute magnitudes for galaxies based on the TF distances to their parent groups, is that this may result in spurious KIII candidates in situations where background galaxies have been incorrectly assigned to foreground groups. This leads to underestimated galaxy distances and therefore absolute magnitudes that are much too faint. A potential indication of this is when the redshifts of faint outliers appear to be significantly higher than those of the groups hosting them. Peculiar motions may account for some of these offsets, but for four cases in Table~\ref{KIII_candidates} (objects 1, 2, 9 and 10), the differences between the observed galaxy redshifts $cz_\mathrm{obs}$ and the group distances $r_{\mathrm{grp,M}}$ correspond to velocity offsets of more than 3000 km s$^{-1}$, which seems extremely high. However, scrutiny of the SFI++ catalog entries reveals that the observed redshifts of these four galaxies agree well with the average group redshifts, and that it is the group TF distances $r_{\mathrm{grp,M}}$ themselves that are dubious. These four outliers belong to two small groups with only three TF members each. In both cases, the third member happens to be a low-mass galaxy ($\log_{10}W_\mathrm{TF} \approx 2$) with very high luminosity ($\approx 3$ magnitudes above the TF relation). Hence, the inferred TF distances to these groups seem to be biased by small number statistics and TF outliers, albeit outliers in the opposite sense of what we are looking for. 

Since the adopted distances to the groups hosting our KIII candidates are sensitive to the number of other TF galaxies in these structures, groups with many TF members are in general expected to have more reliable distance estimates than those with few members. The 11 outliers in Table~\ref{KIII_candidates} are part of 7 groups, where the smallest ones are the groups with SFI++ identifiers 40702, 30590 and 31043 (3 members each in the SFI++ catalog) and group 40032, (4 members). The largest groups are group 30654 with 29 members, group 1736 with 10 and group 3407 with a total of 8 members. Here, we have assigned a default class 'B' to candidates with only 3 TF members, as these are more prone to errors, and class 'A' to groups with $\geq 4$ members. Because of the clear-cut problems with the groups hosting objects 1, 2, 9 and 10, these are further degraded from class 'B' to 'C', and are deemed to be the least likely KIII candidates in Table~\ref{KIII_candidates}. When comparing the offsets between observed galaxy redshifts and the mean redshifts of their parent groups, objects 5 and 8 also stand out, with velocity offsets of $\approx 500$ km s$^{-1}$ and $\approx 700$ km s$^{-1}$ respectively. Since these are members of the two largest groups, with SFI++ velocity dispersions similar to or larger than these offsets, this is by itself not a source of concern. However, in the case of object 8, which would belong to class A according to our criteria, we suspect that there is something wrong with the data or the catalog entry, since \citet{Springob et al. 07} lists an apparent magnitude corrected for inclination and extinction that is {\it fainter} than the uncorrected observed magnitude, which is unphysical. This objects has therefore also been transfered to class 'B'.

\begin{figure}[t]
\includegraphics[width=84mm]{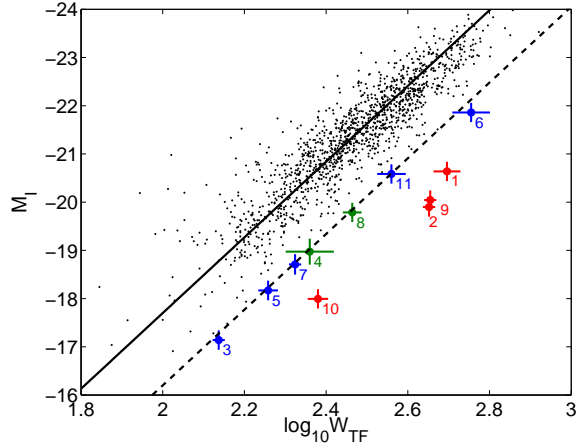}
\caption{Tully-Fisher diagram containing the 1359 objects from \citet{Springob et al. 07} discussed in Sect.~\ref{constraints}, with error bars suppressed for all objects except the KIII candidates to avoid cluttering. The solid line indicates the TF relation and the dashed line the limit below which outliers qualify as KIII candidates according to the \citet{Annis} criterion. Numbered dots mark the 11 outliers discussed in the text, with blue indicating the KIII candidiates of class 'A', green indicating class 'B' and red indicating class 'C'. Some of the errorbars have been slightly exaggerated for plotting purposes.} 
\label{TF_figure}
\end{figure}

Outliers of class A, B and C are marked in Fig.~\ref{TF_figure} as blue, green and red symbols respectively. As seen, the outliers with the least reliable distance estimates (class C) are the ones with the most extreme offsets from the TF relation, whereas objects belonging to class A and B straddle the KIII delimiter line. 

It should be noted that the present analysis is insensitive to extreme KIII objects, for which $\gtrsim 90\%$ of the optical light has been absorbed by Dyson spheres (i.e. galaxies located $\geq 2.5$ mag below the TF), since such disks would not make it into the SFI++ sample in the first place.  A number of such ``dark galaxy'' candidates, with very faint or undetected optical counterparts, have already been uncovered in blind HI sureys \citep[e.g.][]{Haynes et al.}. Most of these have very low masses and velocity widths \citep[e.g.][]{Cannon15,Janowiecki15}, and are therefore in the region where the TF already is known to exhibit considerable scatter \citep{Geha et al.}. Whether there are massive dark galaxies remains an open question -- while such detections have been claimed \citep{Minchin05,Minchin07}, it has also been argued that these could simply be tidal HI structures with extended HI line profiles without ordered rotation \citep{Duc & Bournaud}.

\subsection{The prevalence of Kardashev type III civilizations}
\label{statistics}
It is important to realize that a substantial fraction of galaxies in the SFI++ compilation would likely have been excluded from the sample, had they been converted into KIII objects, as this would have rendered them too faint for detection or otherwise altered their properties in ways that would have prevented them from entering the catalog. In fact, the prospects of finding KIII candidates quickly deteriorate as the condition of TF offset $\Delta(M_I)$ is sharpened. The method becomes far less efficient at $\Delta(M_I)=2.5$ mag (90\% of light absorbed) than for $\Delta(M_I)=1.5$ mag (75\% of light absorbed; our default limit), and fails to produce any useful results at $\Delta(M_I)=5.0$ mag (99\% of light absorbed).  

To convert our harvest of seemingly underluminous disk galaxies into a constraint on the prevalence of star-fed KIII civilizations in disk galaxies, we need to quantify the fraction of objects that would {\it not} have been detectable at a given $\Delta(M_I)$. However, this become highly non-trivial because of the inhomogeneous selection criteria used in the \citet{Springob et al. 07} compilation. A casual upper limit can nonetheless be derived from the effective surface brightness limit of the SFI++ sample. By considering the size-flux distribution of objects in the catalog, we find that KIII-colonized galaxies that end up with a mean $I$-band surface brigthness within the outermost detected $I$-band isophote $\bar{\mu_I} \geq 23.5$ mag arcsec$^{-2}$ are likely to be too faint for detection. To estimate the fraction of KIII candidates that would have been excluded for this reason, we assume that KIII colonization causes a uniform dimming of the disk, and hence a rescaling of the entire surface brightness profile while leaving its overall slope unaltered  (see Sect.~\ref{simulations} for a discussion on this). In this case, KIII colonization makes $\bar{\mu_I}$ fainter, but also makes the galaxy appear smaller, since the faintest detectable isophote is reached at a smaller distance from the centre. The latter effect is important, since some of the subsamples of the SFI++ compilation are based on optical size constraints, which may reject small KIII galaxies even if they are sufficiently bright to be observed. 

To simulate these selection effects, we fit exponential disk profiles to all of the 1359 objects in the SFI++ catalog for which the relevant photometric data is available (about 80\% of the objects). We then lower the surface brightness uniformly by a constant factor ($\Delta(M_I)$), derive the radius of the outermost detectable isophote $r_{\mathrm{max},I}$ and the mean surface brightness $\bar{\mu_I}$ within this radius. Based on the apparent surface brightness limits of the sample, and strict size limits for some of the subsamples in the SFI++ compilation, we consider object likely to be rejected from the sample if $\bar{\mu_I} \geq 23.5$ mag arcsec$^{-2}$ or $r_{\mathrm{max},I}< 30$ arcsec. Through this procedure, we find that $\approx 30\%$, or $\approx 400$ objects would likely have remained detectable after 1.5 mag of dimming (75\% of the light lost). Given 11 candidates out of a sample of 400, one may then conservatively conclude that $\lesssim 3\%$ of local disk galaxies are housing star-fed KIII civilizations. These estimates also reveals the limitations of the method -- only 16\%, 9\% and 3\% of the objects in the sample would have been detectable at $\Delta(M_I)$=2.1 mag, 2.5 and 3.25 mag (85\%, 90\% and 95\% of light absorbed lost). At 99\% dimming, no objects would remain detectable.  

For the \citet{Pierce & Tully} sample of disk galaxies used by \citet{Annis}, one may derive a corresponding limit by considering the apparent magnitude limit at which incompleteness effects set in. Following the discussion in \citet{Pierce & Tully}, we adopt a completeness limit at $m_{I\mathrm{c}}\gtrsim 11$, and argue that $\approx 17$ out of the 57 disk galaxies (i.e $\approx 30\%$) used by \citet{Annis} are sufficiently bright to have been included in the sample even if they had been dimmed by 1.5 mag. By using the photometric data from the SFI++ compilation for the subset of the \citet{Annis} disk galaxies (43 objects) that are included in SFI++, we can aslo perform a completion analysis based on surface brightness similar to that used above for our own sample. This suggests that a slightly larger fraction ($\approx 40\%$) of the \citet{Annis} disks may remain detectable after 1.5 mag of dimming. Hence, the non-detection of KIII candidates reported by \citet{Annis} converts into a $\leq $4--$6\%$ upper limit on the prevalence of KIIIs in disk galaxies.  

This means that, due to the existence of the apparent outlier population shown in Fig.~\ref{TF_figure}, our upper limit on the prevalence of star-fed KIII civilizations in disk galaxies is {\it at most} a factor of $\approx 2$ times stronger than that inferred from the original \citet{Annis} study ($\leq 3\%$ for our sample compared to $\leq 4$--$6\%$ for the Annis sample), despite the fact that we are analyzing a disk galaxy sample that is larger by a factor of $\approx 20$. As discussed in Sect.~\ref{reliability}, many of the outliers we find are likely to be due to inaccurate distances, and this is an effect that will plague all large TF surveys -- the \citet{Annis} approach, with focus on just a single, well-defined cluster, provides a cleaner sample in this respect. The presence of a small fraction of disk galaxies that may appear underluminous due to low star formation activity or unusually high dust content just compounds the problem. Applying the TF method to even larger samples may therefore not do much to improve the upper limits on the KIII civilizations, unless auxiliary data allows for stronger constraints on the nature of the outliers.

In Sect.~\ref{nature}, we discuss the use of morphological information and infrared data to further probe the KIII status of our TF outliers. As none of our 11 TF outliers stand out as strong KIII candidates based on these additional tests, we may tentatively conclude that a fraction $\lesssim 1/400$ (i.e. $\lesssim 0.3\%$) of disk galaxies host star-fed KIII civilizations (i.e. a constraint 10--20 times stronger than that of \citealt{Annis}), based on the assumption that none of the 11 candidates are bona fide KIII objects.

\begin{table*}
{\scriptsize
\caption{Candidates for host galaxies of star-fed Kardashev type III civilizations}
\begin{tabular}{llllllllllllll}
\hline
 No. & Ra & Dec & Name & $\log W_\mathrm{TF}$ & $\sigma(\log W_\mathrm{TF})$ &$m_{I\mathrm{c}}$ & $M_I$ & $\sigma(M_I)$  & $r_{\mathrm{grp, M}}$  & $cz_\mathrm{gal}$ & Type & Group & Class \\
    &  (J2000) & (J2000) &      &  (km s$^{-1}$) & (km s$^{-1}$)& & & & (km s$^{-1}$) & (km s$^{-1}$) & & & \\
\hline 
1	&	00 09 48.2 &+27 49 55 &NGC 0022& 2.696 & 0.033  &12.42 & -20.64 & 0.13 & 2862 & 7980 & Sb &40702 & C  \\    
2	&	 00 11 45.1& +28 29 56 &UGC 00108 &  2.652 &  0.007 &13.16 & -19.90 & 0.13 &2862 & 7709 & Sb &40702 & C \\    
3	&	10 01 47.9 &+36 29 56 &UGC 05394 &  2.137 & 0.009  & 14.45& -17.14 & 0.23 & 1455 & 1685 & Sc & 40032 & A \\ 
4	&	11 45 41.2& -28 22 03 &ESO 440-G004 &  2.360 & 0.059  &13.11 &  -18.98 & 0.29 &  1827 & 2182 & Scd &30590 & B \\   
5   &  12 32 03.2 & +16 41 13 &NGC 4502 &  2.258 & 0.024  & 12.89 &-18.17& 0.28 & 1140 & 1944 & Scd &30654 & A  \\ 
6	&	12 48 22.9 & +08 29 15  &NGC 4698 &  2.755 & 0.046  &9.20  & -21.86 & 0.27 &  1140  & 1330 & Sa  & 30654 & A\\
7	&  12 54 48.5 & +19 10 34  &IC 3877 &  2.324 & 0.012   & 12.35 & -18.71 & 0.28 &  1140  & 1216 & Sc & 30654 & A \\
8   & 13 26 12.8 & -27 29 06 & AGC 530433 &  2.464 & 0.023 & 16.21& -19.79 & 0.09 &  11080 & 9992 & S0/a &  1736 & B \\
9  &	22 13 30.6 & -27 33 30  &ESO 467-G023 &  2.655 & 0.013  &11.82 & -20.05 & 0.20 & 1651  & 4976 & Sb &31043 & C \\
10  &	22 15 48.2 & -27 30 44  &ESO 467-G034 &  2.380 & 0.025 &13.87 & -18.00 & 0.21 &  1651  & 4859 & Sb &31043 & C \\ 
11 & 07 03 26.7 & -48 59 40 &AGC 470027 & 2.560 & 0.035 &  15.32 & -20.91 & 0.08 & 12346  & 12574 & Sb & 3407 & A \\ 
\hline
\label{KIII_candidates}
\end{tabular}\\
The entries in the group column correspond to group identifiers from the SFI++ catalog. 
}
\end{table*}

\section{The nature of the Kardashev type III galaxy candidates}
\label{nature}
As described in Sect.~\ref{constraints}, a small fraction of local disk galaxies appear to be significantly underluminous (by a factor of four, i.e. by 1.5 mag) compared to the TF relation.  Individual objects that are TF outliers in this respect have been discussed in the literature before \citep[e.g.][]{Barton et al.,Kannappan et al.,Pizagno et al.}, without much consensus on the likely cause of their low luminosities, although it has been noted that disks of morphological class Sa display a potential systematic offset in this direction of the TF \citep{Kannappan et al.}. As seen in Table~\ref{KIII_candidates}, only 1--2 of the KIII candidates in our final sample bear this classification. In the following, we will discuss auxiliary observations to further probe the nature of our candidates. 

\subsection{Morphological signatures} 
\label{simulations}
As the colonization wave of a civilization in the transition phase between KII and KIII sweeps across a galaxy, conspicuous morphological anomalies may potentially result, but only as long as the Dyson spheres remain confined to specific regions within that object. Once a galaxy is completely colonized, a uniformly dim disk is the more likely outcome. In principle, a completely dark disk may also arise (which would be undetectable with the current method), but this requires that every type of star is targeted for astroengineering, and that the Dyson spheres capture 100\% of the optical light, which may be difficult to achieve in practice.

In the context of Dyson spheres constructed by self-replicating von Neumann probes, colonization will typically engulf a galaxy like the Milky Way in $\lesssim 10^7$--$10^8$ yr \citep[e.g.][]{Tipler80,Valdes & Freitas 80,Barlow13,Nicholson & Forgan 13}, resulting in a uniformly dim system on time scales much shorter than the current ages of large disks ($\sim 10^{10}$ yr). Hence, we are unlikely to catch the system in the KII-KIII transition phase where morphological signatures would be present.

For non-von Neumann type colonizers, estimates on the time required to colonize a galaxy vary greatly depending on assumptions such as vessel propulsion method, number of probes and population dynamics -- from $\sim 10^6$ yr up to $\sim 10^{10}$ yr \citep{Hart,Jones,Newman & Sagan 81,Sagan & Newman 83,Cotta & Morales,Forgan et al.,Wright et al. b}. But even if the time scale for complete colonization is long, would this necessarily result in long-lived morphological anomalies during the colonization phase? In many cases, the answer is no. In the case of very slow-moving colonizing vessels, stellar dispersion would still spread the colonized stars throughout most or all of the galaxy in $\sim 10^8$--$10^9$ yr \citep{Wright et al. b}. A comparison of this time scale with the ages of disks ($\sim 10^{10}$ yr) already indicates that $<10\%$ of KII-KIII transition objects should display large-scale morpholgical anomalies. 

What if the coloinzation efforts were to stop abruptly, once a certain fraction of the stars had been converted into Dyson spheres? In the case of fast-moving vessels (moving faster than the stars within the galaxy), the colonization pattern at the stalling point could be highly irregular, but differential rotation and stellar dispersion would still carry the luminous stars into the colonized regions, and result in a uniformly dim disk in few times the rotational period (typically $\sim 10^8$--$10^9$ yr). However, there are special situations in which morphological curiosities could be long-lasting, due to the limited mixing between the dynamical subcomponents of galaxies. As an extreme case, consider a colonization wave that spreads exclusively throughout the bulge\footnote{Here taken to mean a classical bulge formed through mergers or monolithical collapse, as opposed to a pseudo-bulge which forms from secular processes in the disk.} of a disk galaxy, but stops abruptly once some fraction of bulge stars have been turned into Dyson spheres. This would render the bulge significantly underluminous, and while some of these Dyson spheres may wander into the disk, they will not all end up on disk-like orbits for any appreciable amount of time. After all, the reason why significant age and metallcitiy differences are observed between bulge and disk populations in individual galaxies \citep[e.g][]{Sanchez-Blazquez et al.} is that stars in these subcomponents have retained their orbital identities for several billions of years.

In Fig.~\ref{fig_morphology}, we use a simple numerical model of galaxy colonization to illustrate a long-lasting morphological anomaly that arises because of a stalling colonization wave. In these simulations, the stellar population of a galaxy is represented by $5\times 10^5$ star particles, each representing a large number of stars ($\sim 10^6$). The star particles are distributed throughout the thin disk, thick disk, bulge and halo of a galaxy with dimensions similar to the Milky Way (disk scale lengths 3 kpc; thin disk scale height 0.3 kpc; thick disk scale height 1 kpc; bulge radius 1.5 kpc, halo radius 30 kpc), and assigned velocities depending on which component they belong to. The masses of the bulge, thick disk and stellar halo are assumed to be $30\%$, $5\%$ and $1\%$ of the mass of the thin disk, and the number of particles in these components scale accordingly. The particles within the disks are assigned rotational velocities given by the universal rotation curve for spirals specified in \citet{Persic & Salucci}. An additional velocity component in the polar direction is also assigned to make each particle oscillates around its initial position with a maximum polar velocity of 30 km s$^{-1}$. A random radial velocity component with velocity dispersion 10 km s$^{-1}$ is also included in every time step. The bulge and halo particles are assigned two velocity components: a fixed rotational velocity of 100 km s$^{-1}$ and 200 km s$^{-1}$ respectively in randomly-inclined orbits, plus the same random radial migration velocities (with respect to their inclined orbital planes) as disk particles. 

A random star particle at 8 kpc from the centre (the distance of the Sun from the centre of the Milky Way) is chosen as the starting point of the colonization, and the evolution of the colonization wave is then tracked throughout subsequent timesteps as probes and star particles move about. We adopt a colonization strategy in which each new colony immediately sends out a very large number of probes with velocities $10^{-3}$c (i.e. somewhat faster than the average star particle) in all direction as soon as it is captured. This scheme allows 50\% of the star particles in the model galaxy to turn into colonies in less than $10^8$ yr, after which we assume further colonization attempts to cease. This 50\% level is for plotting purposes chosen as the end stage of the colonization efforts, but the 75\% case that more closely matches our observational search criteria would not produce results that are fundamentally different.

The model galaxy is shown in face-on and edge-on orientations throughout Fig.~\ref{fig_morphology}a-f, in various stages before and after the stalling phase. The colour scheme for the star particles indicates the projected surface mass density at their radius from the centre, with purple/blue indicating the highest stellar density, and red the lowest. Once a star particle is tagged as colonized, it is no longer considered to contribute to the integrated optical luminosity of the galaxy, and the colour of that star particle is then set to black. Fig.~\ref{fig_morphology}a depicts the disk just $5\times 10^6$ yr (at $<1\%$ colonization) after the release of the first colonizing vessels. Fig.~\ref{fig_morphology}b shows the corresponding distribution of colonies at $2\times 10^7$ yr ($\approx 15\%$ colonization). Since the adopted scheme allows for more efficient colonization in regions of high stellar density, the colonization wave is making its way towards the inner regions of the galaxy and has at this point begun overtaking the bulge. The inner regions of the galaxy are then relatively quickly turned into stellar real estate, making the galaxy reach 50\% colonization just a few Myr later (Fig.~\ref{fig_morphology}c). We assume that no more star particles are converted into Dyson spheres beyond that point, and show the state of the same simulated galaxy $2.5 \times 10^8$ yr, $5\times 10^8$ yr and $1\times 10^9$ yr later throughout Fig.~\ref{fig_morphology}d-f. Differential rotation pulls the colonized patch of the disks into a dark spiral, which -- when coupled to random radial migration of stars within the disk-- will eventually turn into a uniformly faint disk. The situation at $2.5\times 10^8$ yr after the stalling point (Fig.~\ref{fig_morphology}d) already closely resembles a system of galactic dust lanes (although without the associated optical reddening), and would not necessarily stand out as anomalous upon morphological inspection of an image in a single filter. Since the simplified simulations presented here do not take density waves, bar instabilities or perturbations due to satellite galaxies into account, it could be argued that the ``dark spiral'' of colonized stars may disperse on a time scale somewhat shorter than suggested by Fig.~\ref{fig_morphology}. However, if a new bulge population does not form (due to mergers or secular evolution of the disk), the central region will remain underluminous for billions of years, since the mixing of particles in the disk and bulge is empirically known to be a very slow process.

To summarize, morphological curiosities are possible, albeit not necessary, signatures for KIII galaxies. Based on optical images of our KIII candidates publicly available through the NASA/IPAC Extragalactic Database (NED), none of them appear to display any conspicuous morphological anomalies. Object 4 (ESO 440-G004) in Table~\ref{KIII_candidates} has a disturbed/irregular appearance, with sprawling spiral arms connected to a central bar \citep{Matthews & Gallagher}, but this is more likely to stem from a merger and does not resemble anything in our colonization simulations. Hence, we find no reason to attribute the optical morphologies of any of our KIII candidates to galaxy-scale astroengineering projects.

\begin{figure*}
\centering
\subfloat{\includegraphics[scale=0.3]{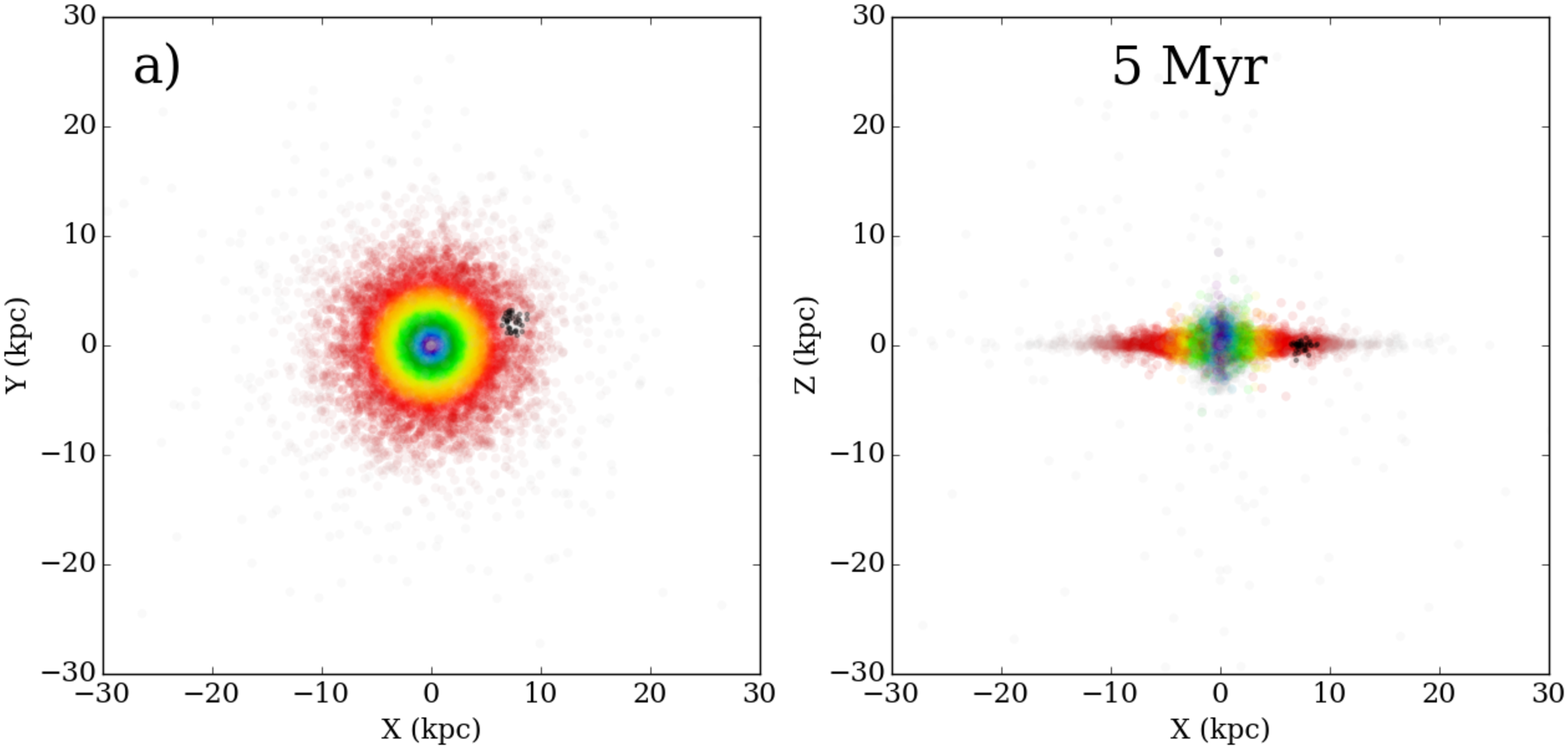} } \,
\subfloat{\includegraphics[scale=0.3]{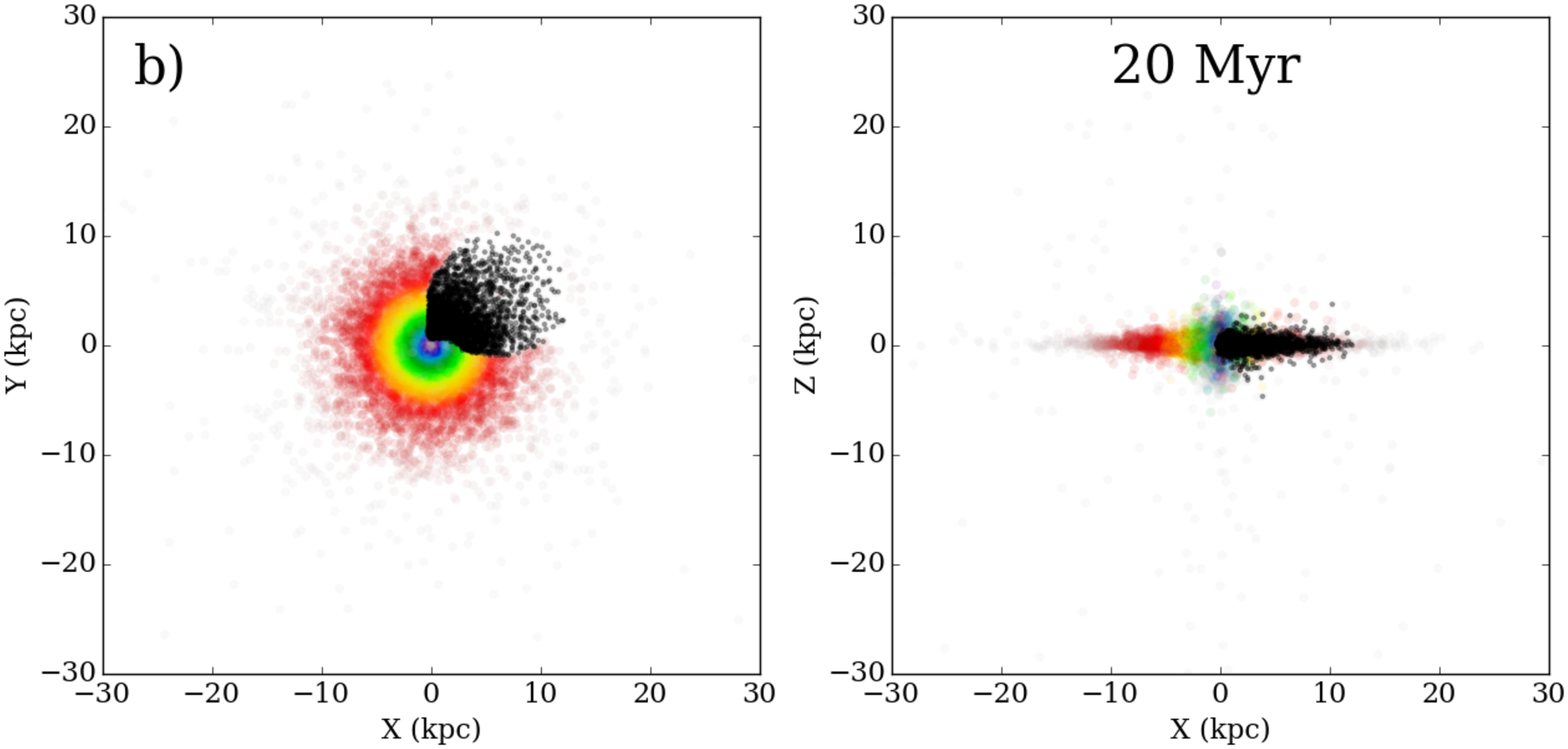} } \,
\subfloat{\includegraphics[scale=0.3]{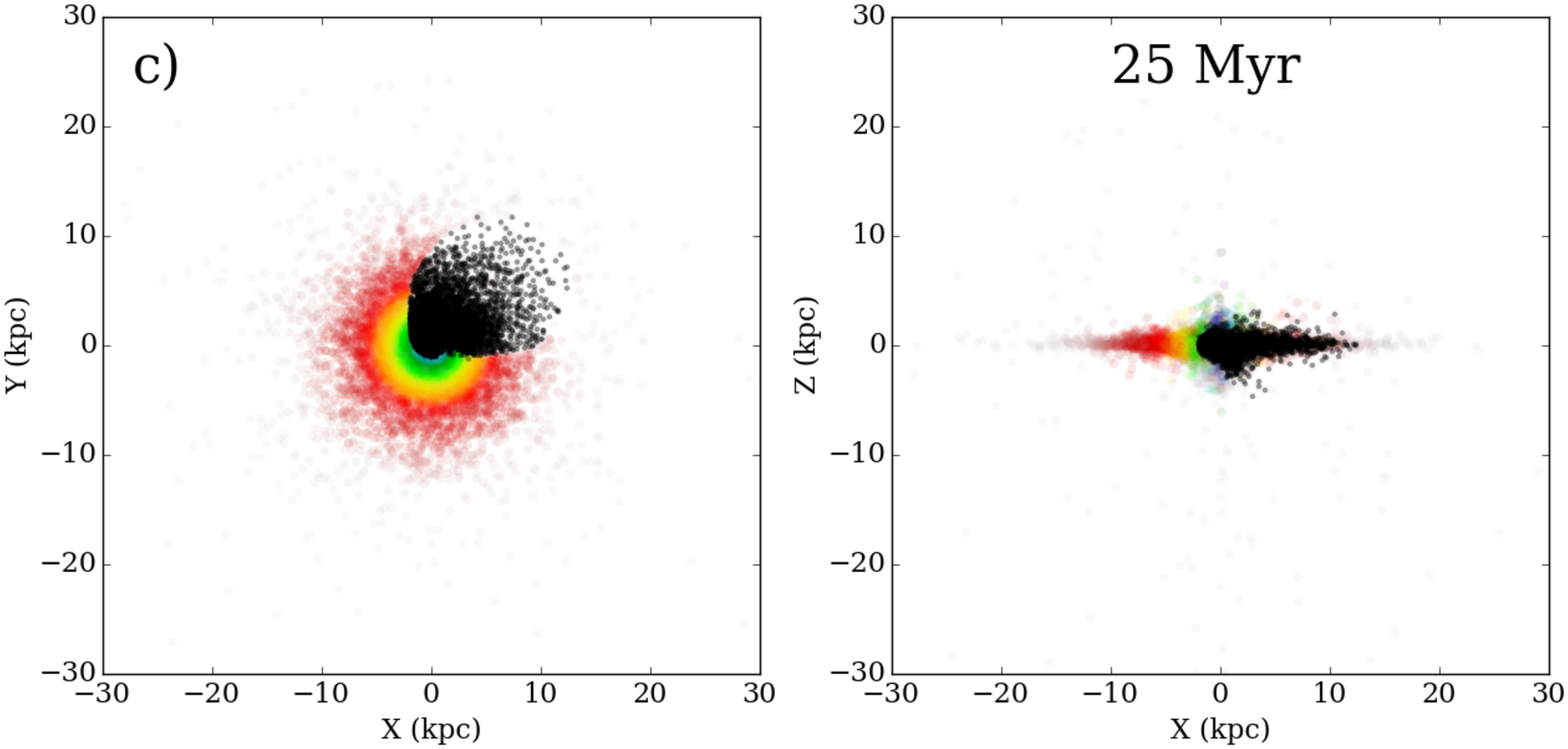} } \,
\caption{Snapshots from a simulation in which star particles within a disk galaxy are colonized by rapidly replicating probes moving at $10^{-3}$ c, but where colonization efforts cease once 50\% of the star particles have been colonized. The purple-red color scale indicates the stellar density in the uncolonized disk, with purple marking the highest density and red the lowest. Colonized particles are marked in black. {\bf a)} The disk in face-on and edge-on orientations 5 Myr after the release of the colonization wave from a position 8 kpc from the centre. {\bf b)} The state of the galaxy at $\approx 15\%$ colonization, after 20 Myr. 
{\bf c)} The state of the galaxy at $\approx 50\%$ colonization, after 25 Myr, at which point no more star particles are assumed to be colonized.  In subsequent panels, the state of the simulated galaxy is shown at $2.5\times 10^8$ yr ({\bf d}), $5\times 10^8$ yr ({\bf e}) and $1\times 10^9$ yr ({\bf f}) after this stalling point. The time stamp in each panel indicates the time since the launch of the first colonizing vessels. While differential rotation and radial migration make the colonized stars in the disk tend towards a uniform disk distribution throughout panels ({\bf e})--({\bf f}), the bulge colonies remain concentrated to the centre, leaving the inner galaxy significantly underluminous at optical wavelengths for billions of years. See main text for further details.}
\label{fig_morphology}
\end{figure*}

\begin{figure*}
\ContinuedFloat
\centering
\subfloat{\includegraphics[scale=0.3]{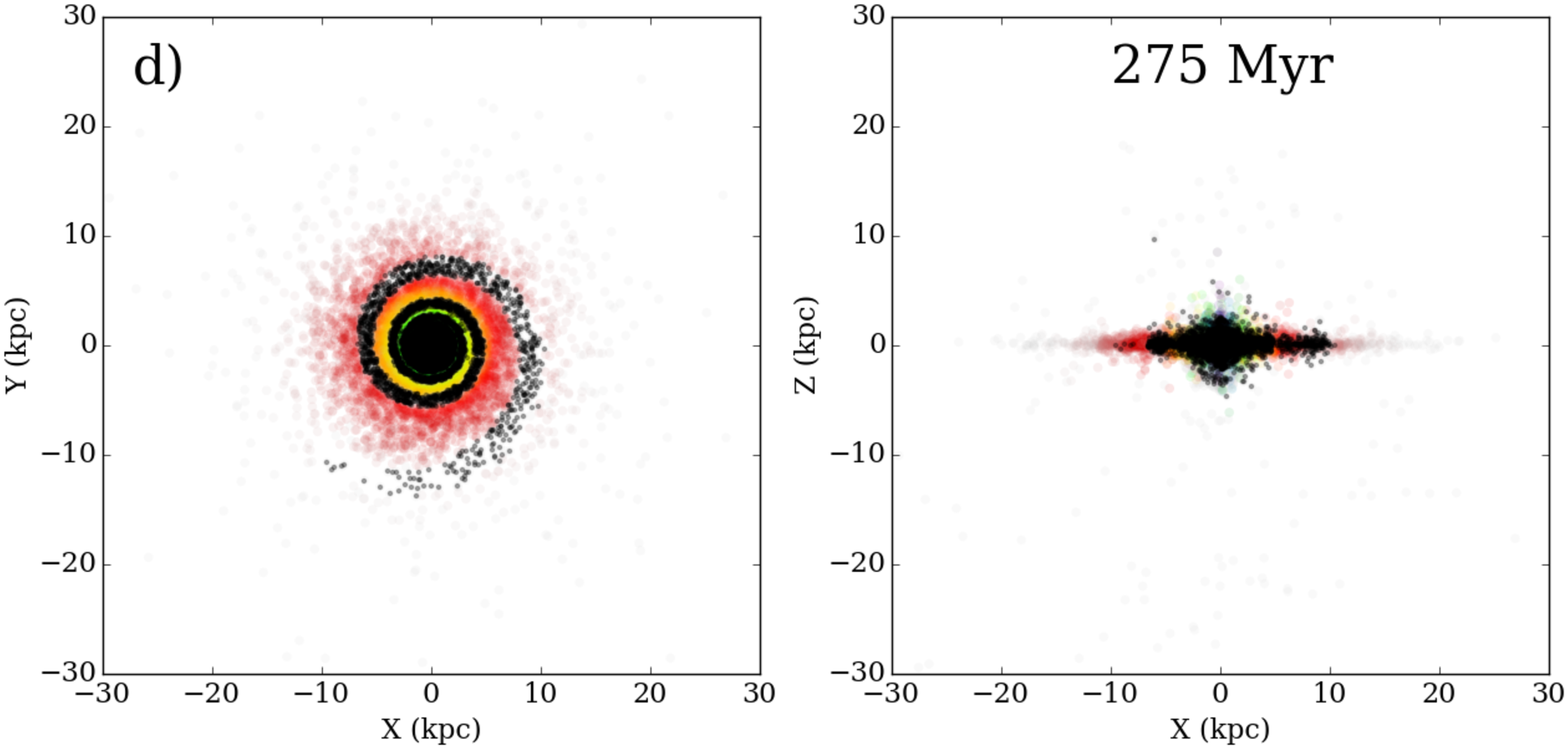} } \\
\subfloat{\includegraphics[scale=0.3]{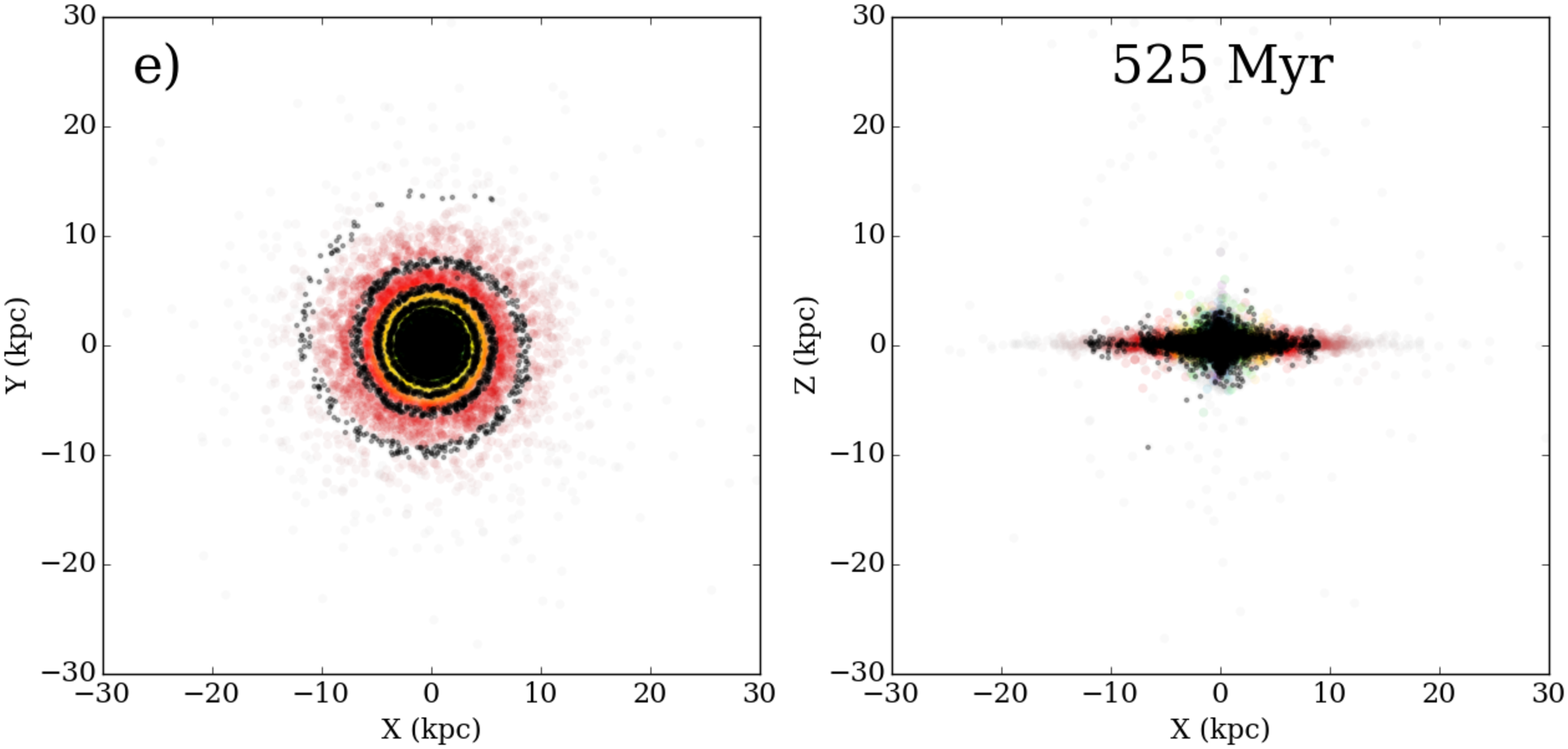} } \\
\subfloat{\includegraphics[scale=0.3]{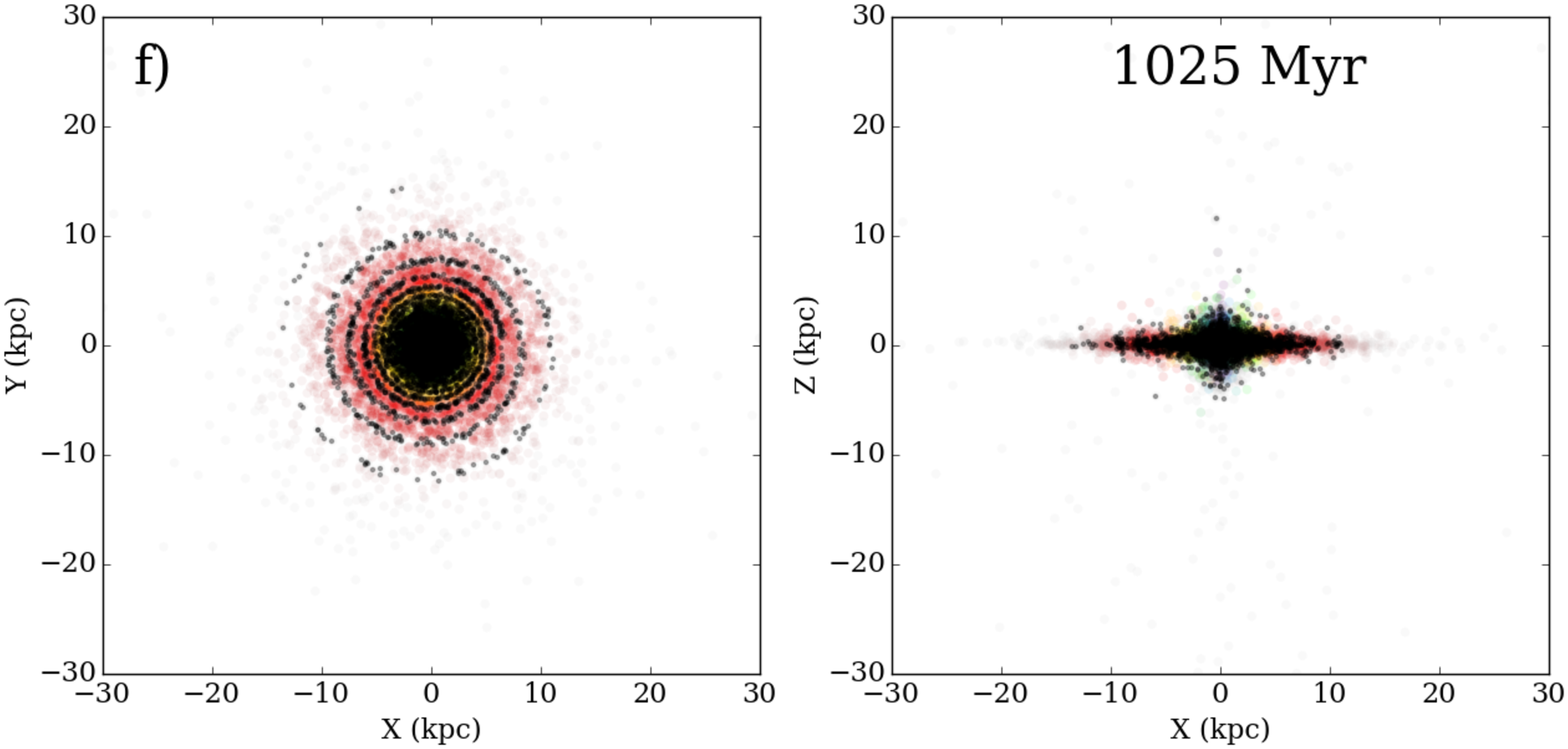} } \,
\caption{Continued} 
\end{figure*}

\subsection{Infrared excess}
While Dyson spheres may be very efficient at absorbing the ultraviolet and optical radiation of stars, waste heat would probably still need to be radiated away. For Dyson spheres constructed out of currently known materials, the structure would need to operate at a temperature that places the peak of the waste radiation at infrared wavelenghts. \citet{Dyson} suggested that one should search for excess radiation at $\approx 10\ \mu$m, which corresponds to the radiation peak of $T\approx 300$ K black body, and a number of searches have already used the InfraRed Astronomical Satellite (IRAS, operating at 12--100 $\mu$m)  to search for Dyson spheres operating at $\approx 50$--600 K \citep{Slysh,Timofeev et al.,Jugaku & Nishimura,Carrigan 09}. The G-HAT survey ({\it Glimpsing Heat from Alien Techonologies}; \citealt{Wright et al. b,Wright et al. c,Griffith et al.}) is currently also looking for such infrared excess signatures (from both KII and KIII civiliziations) using data from the Wide-field Infrared Explorer (WISE).

In the case of galaxies, predicting the exact amount of infrared excess at a specific wavelength turns into a highly non-trivial spectral synthesis problem, since the flux excess depends both on the type of stars targeted by Dysonian astroengineering, on the temperature at which the Dyson sphere operates and -- to some extent -- on the intrinsic dust properties of the galaxy. Our KIII candidates are underluminous by $\gtrsim 1.5$ magnitudes in the $I$-band, but unless Dyson spheres are constructed around {\it all} types of stars, this provides but a poor handle on the bolometric luminosity absorbed by the Dyson spheres (and re-emitted in the infrared). As an example, consider the case where the type of stars that contribute the most to the $I$-band flux are exclusively targeted by Dyson sphere projects. This would correspond to an absorbed  total luminosity {\it lower} than that inferred from reducing the integrated, bolometric stellar luminosity of the entire galaxy by 1.5 mag from its original value. If, on the other hand, the stars dominating the $I$-band are among the least preferred targets of astroengineering, the absorbed fraction may correspond to more than a 1.5 mag reduction of the bolometric, integrated stellar luminosity of the galaxy. Furthermore, part of the optical flux of star-forming galaxies comes in the form of nebular emission rather than direct starlight \citep[e.g.][]{Zackrisson et al. a,Zackrisson et al. b}. If the hot stars (spectral types O and B) that contribute the most to the photoionization of the interstellar medium are enshrouded in Dyson spheres, the total flux decrease at optical wavelengths will be greater than that inferred by summing up the direct optical flux contribution from these stars. Since much of the dust heating responsible for the mundane mid/far-infrared fluxes of disk galaxies would also decrease as these stars are converted into Dyson spheres, the dust contribution to the mid/far-IR fluxes of KIII galaxies would also be lowered. 

Here, we will not attempt a full-blown spectral synthesis analysis of our KIII candidates, but -- given that a bona fide KIII candidate would be expected to display some sort of IR excess --  simply investigate whether there is something clearly anomalous about the infrared properties of our KIII candidates. While IRAS, Spitzer and ISO data are available for {\it some} of our candidates, all of them have been observed by WISE at 3.4, 4.6, 12 and 22 $\mu$m (although a few are undetected in the 12 and 22 $\mu$m passbands). Here we will focus on the 4.5 and 22 $\mu$m bands, since these are the least affected by emission features from polycyclic aromatic hydrocarbon (PAH) molecules \citep{Wright et al. a}, and should be sensitive to infrared excesses associated with Dyson spheres operating at $T\approx 80$--1200 K under the assumption that these may be treated as single-temperature black bodies. We note, however, that this assumption may break down in cases where there are variations in the characteristic Dyson sphere temperatures among the colonized stars within a galaxy. Single-temperature black bodies may also be poor representations of stars enshrouded by multiple Dyson shells with partial coverage, where each layer is operating at a different temperature\footnote{As in the case of Robert Bradbury's ``Matrioshka brain''}, or so-called stellar engines with temperature differences between the inner and outer surfaces of partial Dyson shells \citep{Badescu95,Badescu00,Badescu06}.

In Table~\ref{IR_excess}, we present the $L_{4.6}/L_I$ and $L_{22}/L_I$ in-band luminosity ratios (in solar units) for our KIII candidates:
\begin{equation}
\frac{L_{4.6/22}}{L_I} = \frac{10^{-0.4(m_{4.6/22} - M_{4.6/22,\odot})}}{10^{-0.4(m_{I\mathrm{c}} - M_{I,\odot})}}, 
\end{equation} 
where $m_{I\mathrm{c}}$ is the $I$-band apparent magnitudes from \citet{Springob et al. 07} corrected for Galactic extinction and inclination (internal dust attenuation), $m_{4.6}$ and $m_{22}$ represent the 4.6 and 22 $\mu$m apparent magnitudes from WISE whereas $M_I$ and $M_{I,\odot}$are the corresponding $I$-band absolute magnitudes of the target galaxy and of the Sun in these passbands ($M_{I,\odot}=4.10$, $M_{4.6,\odot}=3.27$ and $M_{22,\odot}=3.25$ in the Vega system). Missing entries in the $L_{22}/L_I$ column for two of the KIII candidates indicate that these are undetected ($S/N<2$) at 22 $\mu$m. 

\begin{table}
\centering
\caption{IR-to-optical luminosity ratios of the KIII candidate galaxies from Table~\ref{KIII_candidates}}
\begin{tabular}{lll}
\hline
No. &  $L_{4.6}/L_I$ & $L_{22}/L_I$ \\
\hline 
1	& 0.94 & 67 \\    
2	& 1.02 & 43 \\    
3	& 0.24 & 62 \\ 
4	& 0.22 & 46 \\   
5 & 0.35 & 48 \\ 
6	& 0.43 & 2.7 \\
7	& 0.11 &  -\\
8 & 2.7 &  -\\
9 & 1.9 & 690 \\
10 & 0.54 & 91 \\ 
11 & 4.9 & 900 \\ 
\hline
\label{IR_excess}
\end{tabular}\\
Missing entries in the $L_{22}/L_I$ column indicate non-detections ($S/N<2$) in the 22 $\mu$m band. 
\end{table}

As Dyson spheres are expected to lower the $I$-band luminosities and boost the mid/far-infrared luminosities of stars, one naively expects the $L_{4.6}/L_I$ and/or the $L_{22}/L_I$ ratio of a galaxy to grow substantially when a large fraction of its stars are subject this form of astroengineering. Since these ratios are independent of distance, they are not affected by any lingering distance errors in the TF analysis (Sect.~\ref{reliability}). One caveat is that if the Dyson spheres operate at a very low temperature ($\lesssim 70$ K), the excess heat radiation would mostly affect wavelengths longward of 22 $\mu$m, and whatever flux boost these low-temperature Dyson spheres may contribute at 22 $\mu$m could in principle be offset by a drop in the mundane 22 $\mu$m dust radiation as stars that under normal circumstances keep this dust heated are enshrouded in Dyson spheres. The second problem is that the intrinsic scatter among disks in $L_\mathrm{IR}/L_\mathrm{optical}$ luminosity ratios amounts to a factor of $\sim 100$ \citep[e.g.][]{Kennicutt03}, which means that -- without detailed spectral synthesis modelling -- a truly extreme IR excess is required to stand out.

As seen in Table~\ref{IR_excess}, $L_{4.6}/L_I$ varies by a factor of $\approx 40$ and $L_{22}/L_I$ by a factor of $\approx 300$ among our candidates, with the most extreme objects reaching $L_{4.6}/L_I\approx 5\ L_{4.6,\odot}/L_{I,\odot}$ and $L_{22}/L_I\approx 900 \ L_{22,\odot}/L_{I,\odot}$. However, a similar analysis of \citet{Springob et al. 07} disks with more typical $I$-band luminosities with respect to the TF relation indicates that such objects occasionally also attain $L_{4.6}/L_I$ and $L_{22}/L_I$ ratios this high. 

The Kolmogorov-Smirnoff test rejects (at 5\% significance) the hypothesis that the $L_{4.6}/L_I$ and $L_{22}/L_I$ distributions of our KIII candidates is drawn from the same parent population as a random sample of disks from the \citet{Springob et al. 07} catalog. However, this is primarily due to an excess of objects with {\it very low} IR-to-optical flux ratios in our KIII candidate sample ($L_{4.6}/L_I<1 \ L_{4.6,\odot}/L_{I,\odot}$, $L_{22}/L_I<100\ L_{22,\odot}/L_{I,\odot}$), i.e. objects with ratios of a kind {\it not} expected from bona fide KIII host galaxies. 

\begin{figure}
\includegraphics[width=84mm]{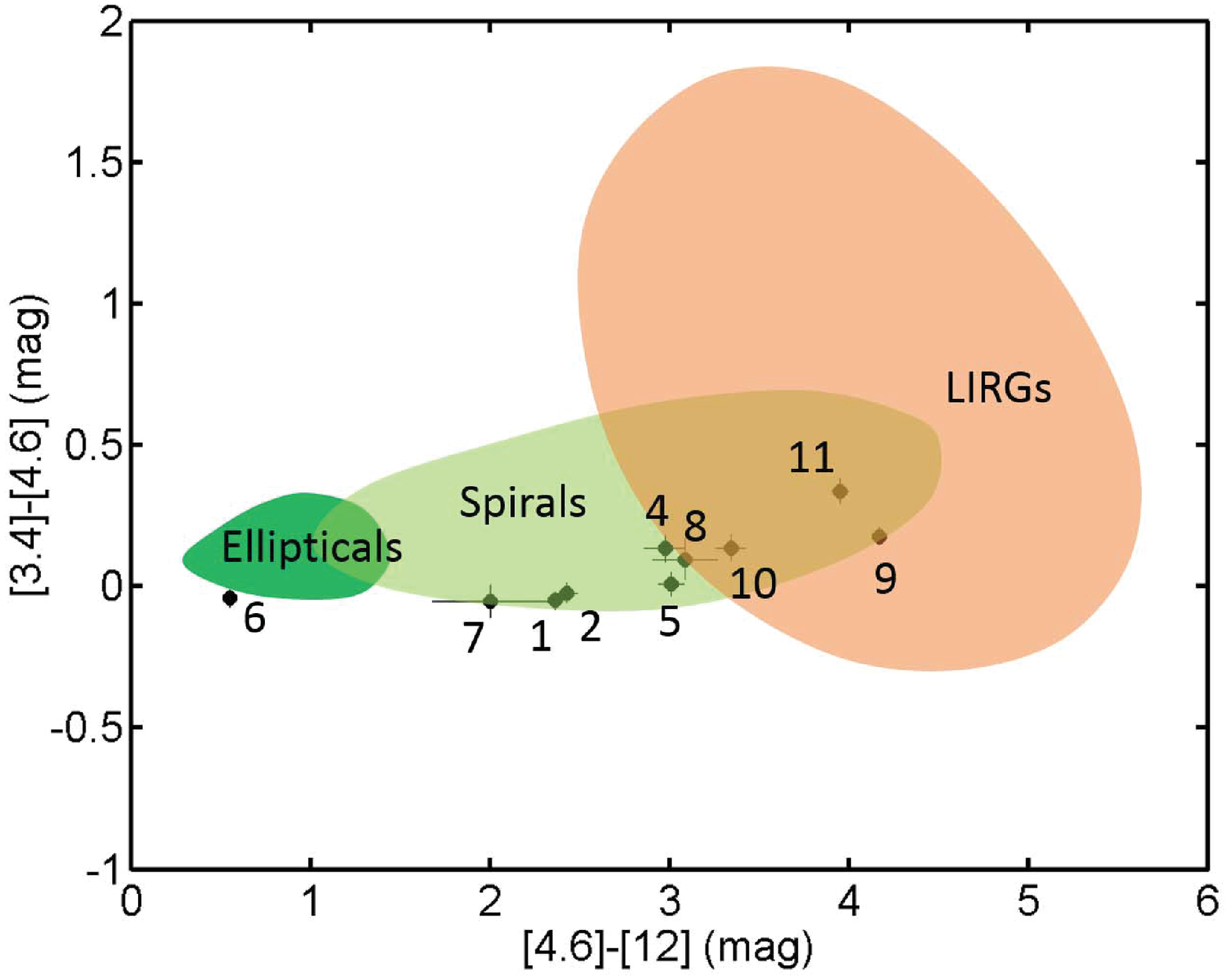}
\caption{WISE $m_{3.4}-m_{4.6}$ vs. $m_{4.6}-m_{12}$ color-color diagram for the 10 of our KIII candidates with WISE detections in these passbands. The objects have been labeled according to their numbers in Table~\ref{KIII_candidates}. The colored patches mark the regions typically inhabited by ellipticals, disks and Luminous Infrared Galaxies (LIRGs) according to \citet{Wright et al. a}. In some cases, the error bars on the measured colors are smaller than the symbols used.} 
\label{IR_fig}
\end{figure}

To explore whether the mid-IR spectra of our KIII candidates shows some other interesting anomalies, we have plotted these objects into the $m_{3.4}-m_{4.6}$ vs. $m_{4.6}-m_{12}$ color-color diagram (Fig.~\ref{IR_fig}), with a few of the diagnostic regions from \citet{Wright et al. a} superposed. Due to a lack of reliable 12 $\mu$m data for one (object 3), only 10 objects are included in this figure.

Most of these KIII candidates fall in the region occupied by disk galaxies, which indicates that their mid-IR flux ratios in these WISE passbands are normal for objects of this type. However, one of the KIII candidates (object 6) is located close to the region typically inhabited by elliptical galaxies, where stellar continuum rather than dust emission dominate the WISE fluxes. Since this object has a fairly low $L_{4.6}/L_I$ ratio and a very low $L_{22}/L_I$ ratio, it is {\it not} a strong KIII candidate based on its IR properties. Instead, a possible explanation for both its anomalously low $I$-band luminosity and lack of pronounced dust emission is that star formation has been quenched in this disk galaxy, thereby effectively shutting off dust heating by massive stars and allowing the Rayleigh-Jeans tail of the stellar radiation to dominate the WISE bands. Indeed, this interpretation is supported by the analysis by \citet{Koopmann & Kenney}, which classify it as ``anemic'' based on the extremely low equivalent width of its H$\alpha$ emission line. Three of the candidates (objects 9, 10 and 11) are located in the part of the spiral galaxy region also occupied by Luminous Infrared Galaxies (LIRGs), and two of these (objects 9 and 11) are also the ones with the highest $L_{22}/L_I$ ratios. Out of these, only object 9 has IR data at longer wavelengths (IRAS; 12--100 $\mu$m). This candidate is not a LIRG according to the standard defintion ($L_\mathrm{IR}>10^{11}\ L_\odot$), but it does have a very high IR-to-optical luminosity ratio for its type (total IR to B-band ratio $L_\mathrm{IR}/L_B \approx 5$ compared to $\approx 0.4$ for a typical Sb galaxy; \citealt{de Jong et al.}). However, other disk galaxies with equally extreme ratios are known \citep{Kennicutt03}.

Despite the complications in predicting the exact SEDs of KIII galaxies, one may still set lower limits on the expected $L_{4.6}/L_I$ and $L_{22}/L_I$ ratios by adopting the very conservative assumption that {\it only} the $I$-band luminosity is absorbed by Dyson spheres, and that this is re-radiated as a single-temperature black body. In the case of black body Dyson spheres of temperature $T=75$, 100, 300 600 K, a 1.5 mag decrement in the $I$-band results in minimum luminosity ratios of $L_{22}/L_I>300$, 800, 850 and 210. For $T=400$, 600 and 1000 K black bodies, $L_{4.6}/L_I>4.7$, 120 and 140, respectively. These lower limits already rule out most of the objects in Table~\ref{IR_excess} as KIII candidates with single-temperature Dyson spheres in the $T>75$ K range. A detailed comparison reveals that $T\approx 400$--1200 K Dyson spheres are ruled out for {\it all} candidates based on the $L_{4.6}/L_I$ ratios, and that 75--800 K Dyson spheres are ruled out for all but objects 9 and 11 based on the $L_{22}/L_I$ ratios.

The data in Table~\ref{IR_excess} would make object 9 consistent with KIII activity under the assumption of Dyson spheres operating at very low temperatures ($T<100$ K). However, since this object belongs to class 'C' due to a dubious distance estimate (see Sect.~\ref{constraints}), we do not considered it a strong KIII candidate. This leaves object 11, which has the highest $L_{22}/L_I$ and $L_{4.6}/L_I$ ratios among our KIII candidates (consistent with KIII activity in the case of $T<400$ K Dyson sphere, under the very conservative assumption that only the I-band flux is absorbed) and has a redshift without any siginficant velocity offset from that of its the parent group.
However, the fact that this object straddles the KIII candidate delimiter line (Fig.~\ref{TF_figure}) makes its candidate status sensitive to the details of how the extinction corrections are made. Our luminosity-dependent extinction corrections differ somewhat from those tabulated in the SFI++ catalog (due to slight differences in the way distances are assigned), and while this difference is in general very small (on average $\approx 0.02$ mag throughout our sample and insignificant for the KIII candidate status of the other 10 objects in Table~\ref{KIII_candidates}), the discrepancy for this particular object happens to be to sufficient to push it $\approx 0.3$ mag closer to the TF if the original SFI++ correction is used. While hardly a strong KIII candidate, this is still the most interesting object uncovered in our search, and could be an interesting target for follow-up observations.

\section{Discussion}
\label{discussion}

Without knowing what an extremely advanced civilization would use the energy collected from large numbers of Dyson spheres for, it is difficult to judge whether this type of astroengineering project would be allowed to grow indefinitely, or cease once some desired energy requirements have been met. If the purpose is to harvest energy to drive supercomputers \citep{Sandberg,Bradbury} the long communication times between distant Dyson spheres may eventually make further expansion of this type of Dysonian colonization unattractive. 

Civilizations that cease their colonization efforts early in the transition stage from Kardashev type II to III would not be singled out using the \citet{Annis} KIII criterion (which requires that 75\% of the optical light has been absorbed), but could potentially still be detectable in nearby, well-resolved galaxies. \citet{Carrigan 10} argue that regions within such galaxies where large numbers of stars have been converted into Dyson spheres would appear as dark patches in high-resolution optical images. As discussion in Sect.~\ref{simulations}, large (kpc-scale) patches of this type would only survive over cosmological time scales under very special circumstances. However, a gravitationally bound star cluster subject to astroengineering could retain its Dysonian signatures for billions of years. Even though optical dimming coupled to an IR excess superficially mimics the effects of dust attenuation, Dysonian astroengineering projects may still give rise to quantitative anomalies in integrated UV/optical/IR colors or spectra of such objects, that would be different from those produced by standard dust attenuation curves. This is especially true if only a subset of stellar types are targeted by astroengineering projects. 

If Dysonian astroengineering ambitions remain firm over time and are not hindered by scale, the colonization wave may in principle spread beyond the boundaries of its galaxy of origin. \citet{Armstrong & Sandberg} outline various schemes through which von Neumann probes may spread to large number of galaxies over cosmological time scales. This provides another opportunity to confirm the status of KIII candidates -- if neighbouring galaxies would also appear anomalously dim, this would constitute supporting evidence for astroengineering. Promising methods to investigate this may be to study the luminosity functions of dwarf galaxies in the vicinity of KIII candidates. In cases where the KIII candidate happens to be located close to another large disk or elliptical  galaxy, a Tully-Fisher or fundamental plane analysis \citep{Annis} would be very interesting. Indeed, several of the KIII candidates in Table~\ref{KIII_candidates} belong to the same galaxy group. Among objects in the 'A' category (indicating that the group has $\geq 4$ members in the catalog and therefore a more reliable TF distance to the group centre), this is the case for objects 5--7. However, these objects also happen to belong to the most nearby group featured in Table~\ref{KIII_candidates} (among the 3\% of groups with the lowest inferred distances in the \citealt{Springob et al. 07} catalog), which makes the distance of individual galaxies from the group centre a significant error source. Without independent (i.e. non TF-based) distance estimates to the individual galaxies in the groups containing several KIII candidates, we are unable to assess how distant these galaxies really are from one another.

\section{Summary}
\label{summary}
By searching a Tully-Fisher sample of $1359$ disk galaxies for the Dysonian astroengineering signatures expected from Kardashev type III civilizations, we are able to set a conservative upper limit of $\lesssim 3\%$ on the fraction of local disks that meet the \citet{Annis} criteria for such civilizations. In this sample, a total of 11 objects are found to be significantly underluminous (by a factor of 4 in the $I$-band) compared to the Tully-Fisher relation, and therefore qualify as Kardashev type III host galaxy candidates according to this test. However, by scrutinizing the optical morphologies and WISE 3.4--22 $\mu$m infrared fluxes of these objects, we find nothing that strongly supports the astroengineering interpretation of their unusually low optical luminosities. Hence, we conclude that their apparent positions in the Tully-Fisher diagram likely have mundane causes, with underestimated distances being the most probable explanation for most of the candidates. Under the assumption that none of them are bona fide KIII objects, we set a tentative upper limit of $\lesssim 0.3\%$ on the fraction of disk galaxies harbouring KIII civilizations.

\acknowledgments
\vspace{5mm}
The authors acknowledge research funding from the Magnus Bergvall Foundation. The anonymous referee is thanked for very insightful comments which helped improve the quality of the paper. This research has made use of the NASA/IPAC Extragalactic Database (NED) which is operated by the Jet Propulsion Laboratory, California Institute of Technology, under contract with the National Aeronautics and Space Administration. This publication also makes use of data products from the Wide-field Infrared Survey Explorer, which is a joint project of the University of California, Los Angeles, and the Jet Propulsion Laboratory/California Institute of Technology, funded by the National Aeronautics and Space Administration


\begin{thebibliography}{}
\bibitem[\protect\citeauthoryear{Annis}{1999}]{Annis}
Annis, J. 1999, JBIS, 52, 33
\bibitem[\protect\citeauthoryear{Armstrong \& Sandberg}{2013}]{Armstrong & Sandberg}
Armstrong, S., Sandberg, A. 2013, Acta Astronautica 89, 1
\bibitem[\protect\citeauthoryear{Badescu}{1995}]{Badescu95}
Badescu, V. 1995, Acta Astronautica 36, 135
\bibitem[\protect\citeauthoryear{Badescu \& Cathcart}{2000}]{Badescu00}
Badescu, V., Cathcart, R. B. 2000, JBIS 53, 297
\bibitem[\protect\citeauthoryear{Badescu \& Cathcart}{2006}]{Badescu06}
Badescu, V., Cathcart, R. B. 2006, Acta Astronautica 58, 119
\bibitem[\protect\citeauthoryear{Barlow}{2013}]{Barlow13}
Barlow, M. T., 2013, IJAsB, 12, 63
\bibitem[\protect\citeauthoryear{Barton et al.}{2001}]{Barton et al.}
Barton, E. J., Geller, M. J., Bromley, B. C., van Zee, L. Kenyon, S. J. 2001, AJ, 121, 625
\bibitem[\protect\citeauthoryear{Beckwith et al.}{2006}]{Beckwith et al.}
Beckwith S.V.W, et al. 2006, AJ, 132, 1729
\bibitem[\protect\citeauthoryear{Bradbury}{2001}]{Bradbury}
Bradbury, R. J. 2001, SPIE, 4273, 56
\bibitem[\protect\citeauthoryear{Bressan et al.}{2012}]{Bressan12}
Bressan, A., Marigo, P., Girardi, L., Salasnich, B., Dal Cero, C., Rubele, S., Nanni, A. 2012, MNRAS, 427, 127
\bibitem[\protect\citeauthoryear{Brin}{1983}]{Brin}
Brin, G. D. 1983, Quarterly Journal of the Royal Astronomical Society, 24, 283 
\bibitem[\protect\citeauthoryear{Cannon et al.}{2015}]{Cannon15}
Cannon, J. M., et al. 2015, AJ, 149, 72
\bibitem[\protect\citeauthoryear{Carrigan}{2009}]{Carrigan 09}
Carrigan, R. A., Jr. 2009, ApJ, 698, 2075
\bibitem[\protect\citeauthoryear{Carrigan}{2010}]{Carrigan 10}
Carrigan, R. A. Jr., 2010, JBIS, 63, 90
\bibitem[\protect\citeauthoryear{Chen et al.}{2014}]{Chen14}
Chen, Y., Girardi, L., Bressan, A., Marigo, P., Barbieri, M., Kong, X. 2014, MNRAS, 444, 2525
\bibitem[\protect\citeauthoryear{Chung et al.}{2002}]{Chung et al.}
Chung, A., van Gorkom, J. H., O'Neil, K., Bothun, G. D. 2002, AJ, 123, 2387
\bibitem[\protect\citeauthoryear{\'Cirkovi\'c \& Bradbury}{2006}]{Cirkovic & Bradbury}
\'Cirkovi\'c M. M., Bradbury, R. J. 2006, New Astronomy, 11, 628
\bibitem[\protect\citeauthoryear{\'Cirkovi\'c}{2009}]{Cirkovic}
\'Cirkovi\'c M. M. 2009, Serbian Astronomical Journal, 178, 1
\bibitem[\protect\citeauthoryear{Cotta \& Morales}{2009}]{Cotta & Morales}
Cotta, C., Morales, A. 2009, JBIS, 62, 82
\bibitem[\protect\citeauthoryear{de Jong et al.}{1984}]{de Jong et al.}
de Jong, T., et al. 1984, ApJ, 278, L67
\bibitem[\protect\citeauthoryear{Duc \& Bournaud}{2008}]{Duc & Bournaud}
Duc, P.-A., Bournaud, F. 2008, ApJ, 673, 787
\bibitem[\protect\citeauthoryear{Dyson}{1960}]{Dyson}
Dyson, F. 1960, Science, 131, 1667
\bibitem[\protect\citeauthoryear{Forgan et al.}{2013}]{Forgan et al.}
Forgan, D. H., Papadogiannakis, S., Kitching, T. 2013, JBIS, 66, 171
\bibitem[\protect\citeauthoryear{Geha et al.}{2006}]{Geha et al.}
Geha, M., Blanton, M. R., Masjedi, M., West, A. A. 2006, ApJ, 653, 240
\bibitem[\protect\citeauthoryear{Giovanelli et al.}{1995}]{Giovanelli et al.} 
Giovanelli, R., et al. 1995, AJ, 110, 1059 
\bibitem[\protect\citeauthoryear{Giovanelli et al.}{1997}]{Giovanelli et al. b} 
Giovanelli, R., Haynes, M. P., Herter, T., Vogt, N. P., Da Costa, L. N., Freudling, W. 1997, AJ, 113, 53 
\bibitem[\protect\citeauthoryear{Griffith et al.}{2015}]{Griffith et al.} 
Griffith, R. L., Wright, J. T., Maldonado, J., Povich, M. S., Sigur${\eth}$sson, S., Mullan, B. 2015, ApJS, 217, 25
\bibitem[\protect\citeauthoryear{Hart}{1975}]{Hart}
Hart, M. H. 1975, QJRAS, 16, 128
\bibitem[\protect\citeauthoryear{Haynes et al.}{2011}]{Haynes et al.}
Haynes, M. P., et al. 2011, AJ, 142, 170
\bibitem[\protect\citeauthoryear{Inoue \& Yokoo}{2011}]{Inoue & Yokoo}
Inoue, M., Yokoo, H. 2011, JBIS, 64, 59
\bibitem[\protect\citeauthoryear{Janowiecki et al.}{2015}]{Janowiecki15}
Janowiecki, S., et al. 2015, ApJ, 801, 96
\bibitem[\protect\citeauthoryear{Jones}{1976}]{Jones}
Jones, E. M. 1976, Icarus, 28, 421
\bibitem[\protect\citeauthoryear{Jugaku \& Nishimura}{2004}]{Jugaku & Nishimura}
Jugaku, J. \& Nishimura, S., 2004, Bioastronomy 2002: Life Among the Stars, Proceedings of IAU Symposium 213, R. Norris and F. Stootman, eds, ASPC 213, 437
\bibitem[\protect\citeauthoryear{Kannappan et al.}{2002}]{Kannappan et al.}
Kannappan, S.J., Fabricant, D. G., Franx, M. 2002, AJ, 123, 2358
\bibitem[\protect\citeauthoryear{Kardashev}{1964}]{Kardashev}
Kardashev, N. S. 1964, Soviet Astronomy, 8, 217
\bibitem[\protect\citeauthoryear{Kennicutt et al.}{2003}]{Kennicutt03}
Kennicutt, R. C., Jr., et al. 2003, PASP, 115, 928
\bibitem[\protect\citeauthoryear{Koopmann \& Kenney}{2004}]{Koopmann & Kenney}
Koopmann, R., Kenney, J. D. P. 2004, ApJ, 613, 851
\bibitem[\protect\citeauthoryear{Maccone}{2012}]{Maccone}
Maccone, C. 2012, Acta Astronautica 78, 109
\bibitem[\protect\citeauthoryear{Masters et al.}{2006}]{Masters et al.}
Masters, K. L., Springob, C. M., Haynes, M. P., Giovanelli, R. 2006, ApJ 653, 861
\bibitem[\protect\citeauthoryear{Matthews \& Gallagher}{1997}]{Matthews & Gallagher}
Matthews, L. D., Gallagher, J. S., III, 1997, AJ, 114, 1899
\bibitem[\protect\citeauthoryear{Matthews et al.}{1998}]{Matthews98}
Matthews, L.D., van Driel, W., Gallagher, J. S., III, 1998, AJ 116, 2196
\bibitem[\protect\citeauthoryear{McGaugh et al.}{1998}]{McGaugh98}
McGaugh, S. S., de Blok, W. J. G. 1998, ApJ, 499, 41
\bibitem[\protect\citeauthoryear{McGaugh et al.}{2000}]{McGaugh00}
McGaugh, S. S., Schombert, J. M., Bothun, G. D., de Blok, W. J. G. 2000, ApJ 533, L99
\bibitem[\protect\citeauthoryear{Minchin et al.}{2005}]{Minchin05}
Minchin, R. 2005, ApJ, 622, L21
\bibitem[\protect\citeauthoryear{Minchin et al.}{2007}]{Minchin07}
Minchin, R. 2007, ApJ, 670, 1056
\bibitem[\protect\citeauthoryear{Newman \& Sagan}{1981}]{Newman & Sagan 81}
Newman, W, I., \& Sagan, C. 1981, Icarus, 46, 293
\bibitem[\protect\citeauthoryear{Nicholson \& Forgan}{2013}]{Nicholson & Forgan 13}
Nicholson, A., Forgan, D. 2013, IJAsB, 12, 337
\bibitem[\protect\citeauthoryear{Persic \& Salucci}{1995}]{Persic & Salucci}
Persic, M., Salucci, P. 1995, ApJ, ApJS 1999, 501
\bibitem[\protect\citeauthoryear{Pierce \& Tully}{1988}]{Pierce & Tully}
Pierce, M., Tully, B. 1988 ApJ, 330, 579 
\bibitem[\protect\citeauthoryear{Pizagno et al.}{2007}]{Pizagno et al.}
Pizagno, J., et al. 2007, AJ, 134, 945
\bibitem[\protect\citeauthoryear{Sandberg}{1999}]{Sandberg}
Sandberg, A. 1999 Journal of Evolution and Technology 5, 1
\bibitem[\protect\citeauthoryear{S\'anchez-Bl\'azquez et al.}{2014}]{Sanchez-Blazquez et al.}
S\'anchez-Bl\'azquez, P., et al. 2014, A\&A, 570, 6
\bibitem[\protect\citeauthoryear{Sagan \& Newman}{1983}]{Sagan & Newman 83}
Sagan, C., \& Newman, W, I. 1983, QJRAS, 24, 113
\bibitem[\protect\citeauthoryear{Schlegel et al.}{1998}]{Schlegel et al.}
Schlegel, D. J., Finkbeiner, D. P., Davis, M. 1998, ApJ, 500, 525
\bibitem[\protect\citeauthoryear{Slysh}{1985}]{Slysh}
Slysh, V. I., 1985,  in The Search for Extraterrestrial Life: Recent Developments, ed. M. Papagiannis, D., Reidel Pub. Co., Boston, Massachusetts (1985), p. 315,
\bibitem[\protect\citeauthoryear{Springob et al.}{2005}]{Springob et al. 05}
Springob, C. M., Haynes, M. P., Giovanelli, R., Kent, B. R. 2005, ApJS, 160, 149 
\bibitem[\protect\citeauthoryear{Springob et al.}{2007}]{Springob et al. 07}
Springob, C. M., Masters, K. L., Haynes, M. P., Giovanelli, R., Marinoni, C. 2007, ApJS, 172, 599
\bibitem[\protect\citeauthoryear{Tang et al.}{2014}]{Tang14}
Tang, J., Bressan, A., Rosenfield, P., Slemer, A., Marigo, P., Girardi, L., Bianchi, L. 2014, MNRAS, 445, 4287
\bibitem[\protect\citeauthoryear{Tarter}{2001}]{Tarter a}
Tarter J. C., 2001, ARA\&A, 39, 511
\bibitem[\protect\citeauthoryear{Tarter}{2007}]{Tarter b}
Tarter J. C, 2007, Highlights of Astronomy, 14, 14
\bibitem[\protect\citeauthoryear{Timofeev}{2000}]{Timofeev et al.}
Timofeev, M. Y., Kardashev, N. S., \& Promyslov, V. G., 2000, Acta Astronautica, 46, 655
\bibitem[\protect\citeauthoryear{Tipler}{1980}]{Tipler80}
Tipler, F. J. 1980, QJRAS, 21, 267
\bibitem[\protect\citeauthoryear{Tully \& Fisher}{1977}]{Tully & Fisher}
Tully, R. B., Fisher, J. R. 1977, A\&A, 54, 661
\bibitem[\protect\citeauthoryear{Valdes \& Freitas}{1980}]{Valdes & Freitas 80}
Valdes, F., Freitas, R. A., Jr. 1980, JBIS, 33, 402
\bibitem[\protect\citeauthoryear{Webb}{2002}]{Webb}
Webb S., 2002,  If the universe is teeming with aliens...where is everybody?: fifty solutions to the Fermi paradox and the problem of extraterrestrial life/ Stephen Webb. New York: Praxis Book/Copernicus Books
\bibitem[\protect\citeauthoryear{Wiley}{2011}]{Wiley}
Wiley, K. B. 2011, arXiv1111.6131
\bibitem[\protect\citeauthoryear{Wright et al.}{2010}]{Wright et al. a}
Wright, E. L., et al. 2010, AJ, 140, 1868
\bibitem[\protect\citeauthoryear{Wright et al.}{2014a}]{Wright et al. b}
Wright, J. T., Mullan, B., Sigur${\eth}$sson, S., Povich, M. S. 2014a, ApJ, 792, 26
\bibitem[\protect\citeauthoryear{Wright et al.}{2014b}]{Wright et al. c}
Wright, J. T., Griffith, R., Sigur${\eth}$sson, S., Povich, M. S., Mullan, B. 2014b, ApJ, 792, 27
\bibitem[\protect\citeauthoryear{Zackrisson et al.}{2001}]{Zackrisson et al. a}
Zackrisson, E., Bergvall, N., Olofsson, K., Siebert, A. 2001, A\&A 375, 814
\bibitem[\protect\citeauthoryear{Zackrisson et al.}{2008}]{Zackrisson et al. b}
Zackrisson, E., Bergvall, N., Leitet, E. 2008, ApJ, 676, L9
\bibitem[\protect\citeauthoryear{Zwaan et al.}{1995}]{Zwaan et al.}
Zwaan, M. A., van der Hulst, J. M., de Blok, W. J. G., McGaugh, S. S. 1995, MNRAS, 273, L35
\end{thebibliography}
\end{document}